\newcommand{\be}{\begin{equation}}\newcommand{\ee}{\end{equation}}
\newcommand{\bea}{\begin{eqnarray}}\newcommand{\eea}{\end{eqnarray}}
\documentstyle[12pt]{article}
\def\a{\alpha}\def\b{\beta}\def\g{\gamma}\def\d{\delta}\def\e{\epsilon}
\def\k{\kappa}\def\l{\lambda}\def\L{\Lambda}\def\s{\sigma}
\def\Th{\Theta}\def\th{\theta}

\newcommand{\p}[1]{(\ref{#1})}
\begin{document}
\renewcommand{\thefootnote}{\fnsymbol{footnote}}
\thispagestyle{empty}
\begin{flushright}
Preprint DFPD 95/TH/08, \\

\end{flushright}

\vspace{0.5cm}
\begin{center}
{\large \bf  Twistor--like superparticles revisited}

\vspace{0.7cm}
{\bf Igor A. Bandos, Alexsey Nurmagambetov,}

\vspace{0.2cm}
{\it Kharkov Institute of Physics and Technology}
{\it 310108, Kharkov,  Ukraine}\\
e-mail:  kfti@kfti.kharkov.ua

\vspace{0.5cm}
\renewcommand{\thefootnote}{\dagger}
{\bf Dmitrij Sorokin\footnote{on leave from Kharkov Institute of
Physics and Technology, Kharkov, 310108, Ukraine.\\e--mail:
sorokin@pd.infn.it}},

\vspace{0.2cm}
{\it Universit\`a Degli Studi Di Padova
Dipartimento Di Fisica ``Galileo Galilei''\\
ed INFN, Sezione Di Padova
Via F. Marzolo, 8, 35131 Padova, Italia}

\vspace{0.3cm}
{\bf and}

\vspace{0.3cm}

{\bf Dmitrij V. Volkov}

\vspace{0.2cm}
{\it Kharkov Institute of Physics and Technology}
{\it 310108, Kharkov,  Ukraine}\\
e-mail:  dvolkov@kfti.kharkov.ua

\vspace{1.5cm}
{\bf Abstract}
\end{center}

\bigskip
We consider a formulation of N=1 D=3,4 and 6 superparticle mechanics,
which is manifestly supersymmetric on the worldline and in the target
superspace. For the construction of the action we use only
geometrical objects that characterize the embedding of the worldline
superspace into the target superspace, such as target superspace
coordinates of the superparticle and twistor components. The action does
not contain the Lagrange multipliers which may cause the problem of
infinite reducible symmetries, and, in fact, is a worldline superfield
generalization of the supertwistor description of superparticle dynamics.

\bigskip
PACS: 11.15-q, 11..17+y
\setcounter{page}1
\renewcommand{\thefootnote}{\arabic{footnote}}
\setcounter{footnote}0
\newpage

\section{Introduction.}
Twistor--like doubly supersymmetric formulations of superparticles
\cite{stv}--\cite{gs92}, superstrings \cite{hsstr,hsstr2,gs93} and
supermembranes \cite{tp93} have attracted considerable attention, in
particular, because of a hope to break through the long--standing problem
of the covariant quantization of these theories.

In the twistor--like approach the infinite reducible fermionic
$\k$--symmetry \cite{al,sig}, which causes the problem of covariant
quantization \cite{gsw}, is replaced by local worldsheet supersymmetry
which is irreducible by definition, and the theory is formulated as a
superfield supergravity theory in a worldsheet superspace embedded
into a space--time
target superspace. Thus, the model of this kind possesses double
supersymmetry.

Earlier doubly supersymmetric dynamical systems (of more general physical
contents) were considered by several groups of authors with the aim to
find a version of the superstring theory where the Neveu-Schwarz-Ramond
and Green--Schwarz formulation would appear as a different choice of
gauge \cite{spinsup,vz,stvz,hsstr}.

Several versions of twistor--like doubly supersymmetric particles and
heterotic strings have been constructed in
 $D=3,4$ and $6$ dimensions of space--time, while in D=10 only one
superfield formulation is known \cite{gs92} and, unfortunately, the latter
itself suffers the infinite reducibility problem of a new local
symmetry \cite{gs92} being crucial for the possibility of eliminating
auxiliary degrees of freedom.
Note, however, that at the component level, when auxiliary
fields were eliminated by gauge fixing and solving for relevant equations
of motion, all remaining local symmetries are irreducible. This also takes
place in a twistor--like Lorentz--harmonic formulation of
super--$p$--branes \cite{bh,bzm},which has been developed in parallel with
the superfield twistor approach, and which is, in fact a component
version of the latter \cite{bpstv,bsv}.

The existence of reducible symmetries in twistor--like formulations of
super--p--branes considered so far is connected with the structure of
the Lagrangian, which is constructed with the use of  Lagrange
multipliers, and, in general, it is difficult to endow the latter with a
reasonable geometrical and physical sense. At the same time their role
in the action is to provide geometrical conditions, which characterize
the properties of the embedding of super--p--brane world surface into target
superspace. The most essential trouble which happens with Lagrange
multipliers is that part of them become propagative fields in the case
of N=2, D=10 superstrings and N=1, D=11 supermembranes \cite{gs93,tp93},
which spoils the physical contents of the theory. In this situation it
seems reasonable to revise the way the superfield formulation of
super--p--branes is constructed by use of well defined geometrical
objects on world supersurface and target superspace, and the twistors
(or harmonics) are among them.

So the motivation of the present paper is to
develop a version of the twistor--like formulation which would be free of
the reducibility problem already at the superfield level, and,
 would look ``twistor--like'' as much as possible. The letter,
as we hope, may allow one to better utilize the powerful twistor
techniques  in the structure of supersymmetric theories.

As an example of such alternative formulation we consider the case of
massless N=1 superparticles \cite{bs}
in D=3,4 and 6 superspace--time, and as a basis
for the doubly supersymmetric generalization we use a (super)twistor
formulation of superparticle dynamics originated from papers by Ferber
\cite{fer} and Shirafuji \cite{shir}.

The superfield
twistor--like models of $N=1$ superparticles in $D=3,
4, 6$ and $10$ considered so far are based on the
doubly supersymmetric
generalization of the following massless bosonic particle action
\cite{stv}:
\begin{equation}\label{1}
S= \int d \tau p_{{m}} ( {\dot x}^{{m}} - \bar\l \g^{{m}} \l ) , \qquad
\end{equation}
where $p_{{m}} $ is the particle momentum and $\l^\a$ is a commuting
spinor variable ensuring the validity of the mass shell condition
$
p_{{m}} p^{{m}} = 0 = {\dot x}_{{m}}  {\dot x}^{{m}}
$
due to the Cartan--Penrose representation
$
{\dot x}^{{m}} = \bar\l \g^{{m}} \l
$
of the light--like vectors in $D=3, 4, 6$ and $10$ space--time dimensions
(m=0,...D-1; $\a$=1,...,2(D-2)).

The straightforward doubly supersymmetric generalization of \p{1} is
\cite{stv,gs92}

\begin{equation}\label{1.2}
S= \int d \tau d^{{D-2}} \eta
P_{{mq}} ( D_{{q}} X^{{m}} - iD_{{q}} \bar{\Theta} \g^{{m}}
\Theta ) , \qquad
\end{equation}
where the number $n = D - 2$ of the local worldline supersymmetries
(q=1,...D-2) is
equal to the number of the $\k$--symmetries in  $D=3, 4, 6$ and $10$,
and is half the number of $\Th$;
$
D_{{q}} = {{\partial} \over {\partial \eta^{{q}} } }+i \eta_q
\partial _{\tau}
$
is an odd supercovariant derivative in a worldline superspace
$(\tau , \eta^{{q}} )$ (
$
\{ D_{{q}}, D_{{p}} \} = 2i \d_{pq}  \partial _{\tau}),
$
and $( X^{{m}}, \Th^{{\a}} )$ are worldline superfields which
parametrize the ``trajectory'' of the superparticle in a target
superspace. Bosonic spinor variables $ \l^{{\a}}_{{q}}$ appear in \p{1.2}
as superpartners of Grassmann coordinates $ \th^{{\a}} = \Th^{{\a}}
\vert_{\eta = 0} $ :

\begin{equation}\label{1.3}
\l^{{\a}}_{{q}} =
D_{{q}} \Theta^{{\a}} (\tau, \eta) \vert_{\eta = 0}
\qquad
\end{equation}

The analysis of the action \p{1.2} \cite{stv,gs92} shows that it describes
a superparticle classically equivalent to the massless $N=1$
Brink--Schwarz superparticle in $D=3, 4, 6$ and $10$ .

As we have already mentioned, in $D=4, 6$ and $10$ the action \p{1.2}
possesses a local symmetry \cite{gs92} under the following transformations
of the Lagrange multiplier $P_{{mq}}$ :
\begin{equation}\label{1.4}
\d P_{{mq}} = D_{{p}} \bar{\Xi}_{{qpr}}\g_mD_{{r}} \Th ,
\qquad
\end{equation}
with
$ \bar{\Xi}^{{\a}}_{{qpr}}$ being symmetric and traceless with respect to
the indices $(p, q, r)$. This symmetry is infinite reducible since
$P_{{mq}}$ is inert  under the transformations \p{1.4} with
\begin{equation}\label{1.5}
\bar{\Xi}^{{\a}}_{{qpr}}
= D_{{s}} \bar{\Xi}^{{\a}}_{{qprs}}
\qquad
\end{equation}
where $ \bar{\Xi}^{{\a}}_{{qprs}}$ is again symmetric and traceless,
and \p{1.5} is trivial if
$
\bar{\Xi}^{{\a}}_{{qprs}}
= D_{{s}} \bar{\Xi}^{{\a}}_{{qprst}}
$
and so on and so far.

The reducibility of the transformations \p{1.4} is akin to the
reducibility of the gauge symmetries of the antisymmetric gauge fields.
It is just the problem of reducible symmetries in these theories that
stimulated further development of the quantization procedure which was
consistently followed for finite reducible symmetries \cite{bv}
to  which the gauge transformations of the antisymmetric bosonic tensor
fields belong. However, the general receipt for dealing with the
infinite reducible symmetries is still unknown (see \cite{infred} and
refs. therein). Thus, one has to avoid this problem one way or another.
In the case under consideration we may try to find another form of the
twistor--like superfield action. To this end let us start with

\section{D=3, N=1, n=1 superparticle.}
The bosonic particle action is chosen
to be \cite{fer,shir}
\begin{equation}\label{5}
S= \int d \tau  \bar\l\g_{{m}}\l{\dot x}^{{m}} ,
\end{equation}

To generalize \p{5} to the doubly supersymmetric case one could naively try
(using (3)) to write down an action in the following form
\begin{equation}\label{6}
S= \int d \tau d \eta
D \Theta_{{\a}} D \Theta_{{\b}} D X^{{\a \b}} , \qquad
\end{equation}
Where
$
X^{{\a \b}} \equiv X^{{m}} \g_{{m}}^{{~\a \b}}.
$

However, the action \p{6} does not describe ordinary $N=1,~D=3$
superparticle.
The reason is that \p{6} is invariant under transformations
$$
\d \Th^{{\a}} = \e_{{1}}^{\a} , \qquad
\d X^{\a \b} = \Th^{\a} \e^{\b}_{{2}} + \Th^{\b} \e^{\a}_{{2}} ,
\qquad
$$
so that the target space is not the usual superspace, but one with
additional $\theta$--translations.

Note that action \p{6} is part of a so called spinning superparticle model
considered several years ago \cite{spinsup,vz,sp}.

To construct a doubly supersymmetric action for an $N=1$
superparticle we have to keep only one target space
supersymmetry. The right action turns out to be as follows \cite{bnsv}
\begin{equation}\label{7}
S= \int d \tau d \eta
\L_\a \L_\b (D X^{{\a \b}} - i D \Theta^{\a} \Theta^{\b}
- i D \Theta^{\b} \Theta^{\a} ) , \qquad
\end{equation}
where $\L_{\a} ( \tau , \eta )$ is a commuting spinor superfield
(compare with \cite{fer,shir}).

In addition to $N=1$ target space supersymmetry and $n=1$ local worldline
supersymmetry
\begin{equation}\label{10}
\d \eta = {i\over 2} D\Xi(\tau, \eta), \qquad
\d \tau = \Xi + {1\over 2} \eta D\Xi , \qquad
\d D = -{1\over 2} \partial \Xi D
\qquad
\end{equation}
the action \p{7} is invariant under bosonic transformations
\begin{equation}\label{9}
\d X^{\a \b} = b(\tau, \eta) \L^{\a} \L^{\b} , \qquad
\d \Th ^{\a} = 0 = \d \L ^\a
\qquad
\end{equation}
and under a superfield  irreducible counterpart of the conventional
fermionic  $\kappa$--symmetry
\begin{equation}\label{8}
\d \Th ^{\a} = \k (\tau, \eta) \L^{\a} , \qquad
\d X^{\a \b} = 2i \d \Th^{\{ \a} \Th^{\b \} } , \qquad
\d \L ^\a = 0 , \qquad
\end{equation}
which resembles the fermionic symmetry of
twistor--like component actions for super--$p$--branes \cite{stv,bzm}
( the braces $\{ ...\}$ denote symmetrization of the indices).

The algebra of the transformations \p{9}, \p{8} is closed.

The equations of motion derived from \p{7} are
\begin{equation}\label{11a}
\Pi^{\a \b} \L_{\b} \equiv
( D X^{\a \b} - 2i D \Th^{\{ \a} \Th^{ \b \} } ) \L_{\b} = 0 ,
\qquad
\end{equation}
\begin{equation}\label{11b}
\L_{\b} D \Th^{\b} = 0 , \qquad
\end{equation}
\begin{equation}\label{11c}
\L_{\{ \a} D \L_{\b \} } = 0 ,
\qquad
\end{equation}

The general solutions to \p{11a} and \p{11b} are, respectively,
\begin{equation}\label{12a}
\Pi^{\a \b} = \Psi (\tau, \eta) \L^{\a} \L^{\b} , \qquad
\qquad
\end{equation}
\begin{equation}\label{12b}
D \Th^{\a} = a(\tau, \eta) \L^{\a} , \qquad
\end{equation}
At the same time, from \p{11c} it follows that
\begin{equation}\label{12c}
D \L ^\a = 0 . \qquad
\end{equation}

On the mass shell \p{12a} -- \p{12c} the fermionic superfield $\Psi$ and
the bosonic superfield $a$ transform under \p{8}, \p{9} and \p{10} as
follows:
\begin{equation}\label{13}
\d \Psi = D b - {1\over 2} \partial_{\tau} \Xi \Psi
- 2i a \k ,  \qquad
\d a = D \k - {1\over 2}\partial_\tau \Xi a ,
\end{equation}
Hence, one can fix a gauge
\begin{equation}\label{14}
\Psi = 0,  \qquad   a = 1 ,
\end{equation}
at which \p{12a} and \p{12b} are reduced, respectively, to
\begin{equation}\label{15a}
\Pi^{\a \b} = 0,  \qquad
\end{equation}
\begin{equation}\label{15b}
D\Th^{\a} = \L^{\a}.
\end{equation}
This gauge
\footnote{ Note, that the gauge choice $a = 0$ in Eq. \p{14} is
inadmissible since  then  from \p{12a}, \p{12b} it would follow that
${d \over d\tau} X^m \vert_{\eta = 0} = 0$,
which is, in general, incompatible with boundary conditions
$X^m (\tau_1) \vert_{\eta = 0} = x_{1}$,
$X^m (\tau_2) \vert_{\eta = 0} = x_{2}$.
}
is conserved  under the $\k$--transformations reduced to
the worldline supersymmetry transformations
\begin{equation}\label{17}
D\k-{1\over 2}\partial_\tau \Xi = D(\k+{i\over 2}D\Xi) = 0.
\end{equation}
As a result the twistor superfield $\L^{\a}$ is expressed
in terms of $D\Th^{\a}$ and does
not carry independent degrees of freedom, and in the gauge \p{14} the
equations for $X^{\a\b}$ and $\Th^\a$ coincide with those in the
conventional twistor--like formulation \p{1.2} \cite{stv,gs92}.

Thus we conclude that the doubly supersymmetric action \p{7} is
classically equivalent to \p{1.2} and describes the massless $N=1$
superparticle.

The relationship between the two actions can be understood using the
following reasoning. It was shown in \cite{ps} that for $n=1$ the action
\p{1.2} is classically equivalent to
\begin{equation}\label{18}
S=\int d\tau d\eta(P_{\a\b}\Pi^{\a\b}-{1\over 2}EP_{\a\b}P^{\a\b})
\end{equation}
due to the existence of the following counterparts of the
transformations \p{8}, \p{9} \cite{ps}
\begin{equation}\label{19}
\d X^{\a\b}=\tilde bP^{\a\b},\qquad \d E=D\tilde b, \qquad\Th^\a=0,
\end{equation}
\begin{equation}\label{20}
\d X^{\a\b}=2i\d\Th^{\{\a}\Th^{\b\}},\qquad \d E=-2i\k_\a D\Th^\a, \qquad
\d\Th^\a=\k_\b P^{\b\a},
\end{equation}
which allow one to put the Grassmann superfield $E(\tau,\eta)$ equal to
zero globally on the worldline superspace. \footnote{Note that in contrast
to \p{8} the transformations of eq.~\p{20} correspond to an infinite
reducible $\k$--symmetry \cite{al,sig,gsw}.}

At the same time the variation of \p{18} with respect to $E(\tau,\eta)$
leads to the equation
\begin{equation}\label{21}
P^{\a\b}P_{\a\b}=0,
\end{equation}
which can be solved as
\begin{equation}\label{22}
P_{\a\b}=\L_\a\L_\b
\end{equation}
with $\L_\a$ being an arbitrary bosonic spinor superfield. Substituting
\p{22} into \p{18} we obtain the action \p{7}.

Thus, we have constructed a version of the twistor--like
formulation of the massless $N=1$, $D=3$ superparticle based on eq.~\p{7}
with all symmetries of the model being irreducible. Action \p{7} looks
very much like a worldline superfield generalization of the supertwistor
action by Ferber \cite{fer}.

One can even rewrite \p{7} in a complete supertwistor form
\cite{shir,ced} by introducing the second bosonic spinor
component and the Grassmann component of the supertwistor \cite{fer}:
\begin{equation}\label{28}
M^{\a} =
X^{\a \b} \L_{\b}, \qquad \Upsilon = \Theta^{\a} \L_{\a}. \qquad
\end{equation}
Then, with taking into account the constraints \p{28}, the action (8)
takes the form
$$
S= \int d \tau d \eta (\L_{\a} D M^{\a}
- D \L_{\a} M^{\a}
- 2i \Upsilon D \Upsilon ).
$$

\section{ D=4, N=1 superparticles with n=2 SUSY on the worldline.}

In contrast to the superparticle formulation
based on the action \p{1.2}, the straightforward generalization of the D=3
action \p{8} to the cases D=4,6,10 seems not possible. Indeed, the measure of
integration in these latter cases is even and since the
$\Pi_{q}^{m}\equiv~D_{q}X^{m}-iD_{q}\Theta\gamma^{m}\Theta$ is odd
$P_{m}^{q}$ must be odd as well. Thus, the
Lagrange multiplier can not be replaced by the bilinear combination of
bosonic spinor superfields.

But there is another formulation of D=4
superparticle theory \cite{stv,sp,ds} in terms of worldline chiral
superfields
${X_{L,R}}^{\a\dot\b}=X_{L,R}^m\s_m^{\a\dot\b}$
and $\Theta^{\a}$, $\bar{\Theta}^{\dot\a}$.

It is based on the  concept of double analyticity
\cite{hsstr2,ds}, which means
that the coordinates of the ``analytical" and ``antianalytical" subspaces
$({X_{L}}^{m},\Theta^{\a})$ and $({X_{R}}^{m},\bar{\Theta}^{\dot\a})$ of
the target superspace
$(X^{\a\dot\b}=1/2({X_{L}}^{\a\dot\b}+{X_{R}}^{\a\dot\b}),\Theta^{\a},\bar
{\Theta}^{\dot\a})$ \cite{13sp} should be considered as the worldline
chiral superfields (the bar denotes complex conjugation).
\begin{equation}\label{29}
{X_{L}}^{\a\dot\b}={X_{L}}^{\a\dot\b}(\tau_{L}=\tau+i\eta\bar{\eta},\eta),
\ \ \ \ \ \Theta^{\a}=\Theta^{\a}(\tau_{L},\eta);
\end{equation}
\begin{equation}\label{30}
{X_{R}}^{\a\dot\b}={X_{R}}^{\a\dot\b}(\tau_{R}
=\tau-i\eta\bar{\eta},\bar{\eta}),
\ \ \ \ \
\bar{\Theta}^{\dot\a}=\bar{\Theta}^{\dot\a}(\tau_{R},\bar{\eta}),
\end{equation}
then
\begin{equation}\label{31}
D{X_{R}}^{\a\dot\b}=0=D\bar{\Theta}^{\dot\alpha},\ \ \ \ \ \
\bar{\mbox{\raisebox{0ex}[1.75ex][0ex]{$D$}}}{X_{L}}^{\a\dot\b}=0=
\bar{\mbox{\raisebox{0ex}[1.75ex][0ex]{$D$}}}\Theta^{\alpha},
\end{equation}

where  $D_{q}=(D,\bar{D})$
$$D={\partial\over{\partial{\eta}}}+i\bar{\eta}\partial_{\tau}=
\overline{(\bar{\mbox{\raisebox{0ex}[1.75ex][0ex]{$D$}}})}$$
are the worldline Grassmann derivatives.
In the central basis of the target superspace the superparticle
coordinates  will
satisfy the geometrodynamical condition
\begin{equation}\label{32}
\Pi_{q}^{\alpha\dot\alpha}\equiv
D_{q}X^{\alpha\dot\alpha}-i(D_{q}\Theta^{\alpha})\bar{\Theta}^
{\dot\alpha}-i(D_{q}\bar{\Theta}^{\dot\alpha})\Theta^{\alpha}=0
\end{equation}
if the embedding of the worldline superspace into the target superspace
is defined by the chirality condition (and vice versa)
\begin{equation}\label{33}
Y^{\alpha\dot{\beta}}\equiv {i\over2}({X_{L}}^{\alpha\dot{\beta}}-
{X_{R}}^{\alpha\dot{\beta}})-\Theta^{\alpha}{\bar{\Theta}}^{\dot{\beta}}=0.
\end{equation}

Equation \p{33} can be obtained from the action
with a {\it bosonic} Lagrange multiplier
\begin{equation}\label{34}
S=\int\,d\tau\,d\eta\,d{\bar{\eta}}\,P_{\a\dot\b}
({i\over2}({X_{L}}^{\alpha\dot{\beta}}-
{X_{R}}^{\alpha\dot{\beta}})-
\Theta^{\alpha}{\bar{\Theta}}^{\dot{\beta}}).
\end{equation}

Instead of \p{34} we consider
\begin{equation}\label{35}
S=\int\,d\tau\,d\eta\,d{\bar{\eta}}\,\Lambda_{\alpha}
{\bar{\Lambda}}_{\dot{\beta}}({i\over2}
({X_{L}}^{\alpha\dot{\beta}}-
{X_{R}}^{\alpha\dot{\beta}})-
\Theta^{\alpha}{\bar{\Theta}}^{\dot{\beta}}),
\end{equation}
where $\Lambda_{\alpha}(\tau,\eta,\bar{\eta})$ and
$\bar{\Lambda}_{\dot{\alpha}}(\tau,\eta,\bar{\eta})$ are commuting
spinor superfields.

The action \p{35} possesses N=1 target space supersymmetry, n=2 local
worldline supersymmetry
$$\delta\eta={i\over2}{\bar{\mbox{\raisebox{0ex}[1.75ex][0ex]{$D$}}}}
\Xi(\tau_{L},\eta,\bar{\eta}),
\ \ \ \ \ \delta\tau_{L}=\Xi+\bar{\eta}
\bar{\mbox{\raisebox{0ex}[1.75ex][0ex]{$D$}}}\Xi$$,
\begin{equation}\label{36}
{\delta}D=-{1\over2}(\partial_{\tau}\Xi)D+{i\over4}[D,\bar{\mbox{\raisebox
{0ex}[1.75ex][0ex]{$D$}}}]{\Xi}D,
\end{equation}
U(1) symmetry
$$\delta\Lambda^{\alpha}=i\varphi\Lambda^{\alpha},$$
and also has two additional symmetries analogous to  \p{9} and \p{8}:
\begin{equation}\label{37}
{\delta}{X_{L}}^{\alpha\dot{\beta}}=\bar{\mbox{\raisebox{0ex}[1.75ex][0ex]
{$D$}}}(b(\tau,\eta,\bar{\eta})\Lambda^{\alpha}\bar{\Lambda}^{\dot{\beta}}),
\ \ \ \ \ \ {\delta}{X_{R}}^{\alpha\dot{\beta}}=\overline{({\delta}{X_{L}}^
{\alpha\dot{\beta}})},
\end{equation}
$$\delta\Lambda^{\alpha}=0=\delta\Theta^{\a},\ \ \
\ \ \ \
\delta\bar{\Lambda}^{\dot{\alpha}}=0=\delta\bar{\Theta}^{\dot{\a}},
$$
and
\begin{equation}\label{38}
\delta{X_{L}}^{\alpha\dot{\beta}}=\bar{\mbox{\raisebox{0ex}[1.75ex][0ex]
{$D$}}}(\bar{\kappa}(\tau,\eta,\bar{\eta})
\Lambda^{\alpha}\bar{\Theta}^{\dot{\beta}}),
\ \ \ \ \
\delta{X_{R}}^{\alpha\dot{\beta}}=D(\kappa(\tau,\eta,\bar{\eta})
\Theta^{\alpha}\bar{\Lambda}^{\dot{\beta}}),
\end{equation}
$$\delta\Theta^{\alpha}={i\over2}\bar{\mbox{\raisebox{0ex}[1.75ex][0ex]{$D$}}}
(\bar{\kappa}\Lambda^{\alpha}),
\ \ \ \ \
\delta\bar{\Theta}^{\dot{\alpha}}=-{i\over2}D(\kappa\bar{\Lambda}^{\dot
{\alpha}}).$$

In contrast to the D=3 case both, $b(\tau,\eta,\bar{\eta})$ and
$\kappa(\tau,\eta,\bar{\eta})$, are complex bosonic superfields
(compare with \cite{ps}).

To vary the action \p{35} with respect to the superfields
$X_{L,R}^{m}$, $\Theta^{\a}$, $\bar{\Theta}^{\dot\a}$, the chirality
condition \p{31} should be taken into account.
The simplest way to do this is to
solve \p{31} using the nilpotency of the
covariant spinor derivatives ($DD=0=
\bar{\mbox{\raisebox{0ex}[1.75ex][0ex]{$D$}}}
\bar{\mbox{\raisebox{0ex}[1.75ex][0ex]{$D$}}}$), i.e. to make use of the
representation
$$X_{L}=
\bar{\mbox{\raisebox{0ex}[1.75ex][0ex]{$D$}}}\bar{\psi},\ \ \ \ \ \
X_{R}=D\psi$$,
\begin{equation}\label{39}
\Theta^{\alpha}=
\bar{\mbox{\raisebox{0ex}[1.75ex][0ex]{$D$}}}t^{\alpha},\ \ \ \ \ \
{\bar{\Theta}}^{\dot{\alpha}}=D\bar{t}^{\dot{\alpha}},
\end{equation}
and then vary with respect to unrestricted (``prepotential") superfields
$\bar{\psi}$, $\psi$, $t$, $\bar{t}$.

The equations of motion of $X_{L}$ and $X_{R}$
\begin{equation}
\bar{\mbox{\raisebox{0ex}[1.75ex][0ex]{$D$}}}
(\Lambda_{\alpha}\bar{\Lambda}_{\dot{\beta}})=0=
D(\Lambda_{\alpha}\bar{\Lambda}_{\dot{\beta}})
\end{equation}
read that the real light--like vector
$\Lambda_{\alpha}\bar{\Lambda}_{\dot{\alpha}}$ is constant
\footnote{ This constant light--like vector is just the momentum of the
massless superparticle}.

Fixing the gauge with respect to the U(1) transformations, we get
\begin{equation}\label{41}
\Lambda_{\alpha}=const,\ \ \ \ \ \
{\bar\Lambda}_{\dot\alpha}=const.
\end{equation}

Taking into account \p{41}, the equations of motion of $\Theta^{\alpha}$ and
${\bar\Theta}^{\dot\alpha}$
$$ \delta{S}/{\delta}t^{\alpha}=0,\ \ \ \ \ \ \ \  \delta{S}/
{\delta}\bar{t}^{\dot\alpha}=0$$
are reduced to
\begin{equation}
\Lambda_{\alpha}\bar{\Lambda}_{\dot{\beta}}
\bar{\mbox{\raisebox{0ex}[1.75ex][0ex]{$D$}}}
\bar{\Theta}^{\dot{\beta}}=0,\ \ \ \ \ \
\Lambda_{\alpha}\bar{\Lambda}_{\dot{\beta}}D\Theta^{\alpha}=0
\end{equation}
and, in an assumption that the components of $\Lambda_{\alpha}$ are not
to be zero simultaneously (which is a
convention of the twistor approach), result in
\begin{equation}\label{43}
\bar{\Lambda}_{\dot{\alpha}}
\bar{\mbox{\raisebox{0ex}[1.75ex][0ex]{$D$}}}
\bar{\Theta}^{\dot{\alpha}}=0,\ \ \ \ \ \
\Lambda_{\alpha}D\Theta^{\alpha}=0.
\end{equation}

Much simpler is the derivation of the equations
$$\delta{S}/{\delta{\Lambda_{\alpha}}}=0=\delta{S}/{\delta
{\bar\Lambda_{\dot\alpha}}},$$
which have the form
\begin{equation}\label{44}
\bar{\Lambda}_{\dot{\beta}}Y^{\alpha\dot{\beta}}=0,\ \ \ \ \ \
\Lambda_{\alpha}Y^{\alpha\dot{\beta}}=0.
\end{equation}

The general solution to eqs. \p{43} and \p{44} is
\begin{equation}\label{46}
D\Theta_{\beta}=\bar{a}(\tau,\eta,\bar{\eta})\Lambda_{\beta},\qquad
\bar{\mbox{\raisebox{0ex}[1.75ex][0ex]{$D$}}}
\bar{\Theta}_{\dot{\beta}}=a(\tau,\eta,\bar{\eta})\bar{\Lambda}_
{\dot{\beta}},
\end{equation}
\begin{equation}\label{45}
Y_{\alpha\dot{\beta}}=\xi(\tau,\eta,\bar{\eta})\Lambda_{\alpha}
\bar{\Lambda}_{\dot{\beta}},
\end{equation}
where $a$($\bar{a}$) are (anti)chiral parameters and $\xi$ is real.

On the mass shell the bosonic superfields $\xi$, $a$ transform under
\p{36}, \p{37} and \p{38} as follows:
\begin{equation}\label{48}
\delta{\xi}={i\over2}(b-\bar{b})+(-{1\over2}\partial_{\tau}\Xi+{i\over4}
[D,\bar{\mbox{\raisebox{0ex}[1.75ex][0ex]{$D$}}}
]\Xi)\xi,
\end{equation}
$$
\delta{a}={i\over2}\bar{\mbox{\raisebox{0ex}[1.75ex][0ex]{$D$}}}
{D}\kappa+(-{1\over2}\partial_{\tau}\Xi+
{1\over4}[D,\bar{\mbox{\raisebox{0ex}[1.75ex][0ex]{$D$}}}
]\Xi)a,
$$
which allows one to fix a gauge in the following form
\begin{equation}\label{49}
\xi=0,\ \ \ \ \ \ \ \ \ a=1.
\end{equation}
As in the case of D=3 superparticle the gauge fixing $a=0$ is incompatible
with boundary conditions (see the footnote 2 in section 3).

In the gauge \p{49} the equations of motion are reduced to  \p{33}
(which is the geometrodynamical condition that appeared in a standard
doubly supersymmetric formulation of the D=4
superparticle \cite{stv,ds}), and
\begin{equation}\label{51}
D\Theta_{\beta}=\Lambda_{\beta}
\qquad
\bar{\mbox{\raisebox{0ex}[1.75ex][0ex]{$D$}}}
\bar{\Theta}_{\dot{\beta}}=\bar{\Lambda}_{\dot{\beta}}.
\end{equation}

  Eqs. \p{51} identify $\Lambda_{\beta}$
($\bar\Lambda_{\dot\beta}$) with $D\Theta_{\beta}$
($\bar{D}\bar\Theta_{\dot\beta}$).

In a completely supertwistor form the action \p{35} looks as follows
$$
S=\int d\tau d\eta d\bar\eta(\L_\a M^\a-\bar\L_{\dot\a}\bar M^{\dot\a}
-\Upsilon\bar\Upsilon),
$$
where the supertwistor components $\L_\a,~M^\a,~\Upsilon$ are
constrained to be
$$
M^\a={i\over 2}X^{\a\dot\b}_L\bar\L_{\dot\b},\qquad
\Upsilon=\L_\a M^\a.
$$

\section{ n=4 superfield formulation of D=6 N=1 superparticles.}

As in the case of D=4, there is a  description of D=6, N=1 superparticle
\cite{ds},
 based on the double analyticity concept. An analytic superspace
\begin{equation}\label{53}
(X^{m}_{A}=X^{m}+i\Theta^{i}\gamma^{m}\Theta^{j}w^{+}_{i}w^{-}_{j},
\Theta^{+}_{\alpha}=\Theta^{i}_{\alpha}w^{+}_{i},w^{\pm}_{i})
\end{equation}
is chosen in the D=6, N=1 harmonic
superspace
\begin{equation}
(X^{m},\Theta_{\alpha}^{i},w^{\pm}_{i}),
\end{equation}
where $w^{\pm}_{i}$ are the harmonic variables which parametrize the coset
$SU(2)/U(1)\simeq CP^{1}\simeq S^{2}$ \cite{gikos} (i=1,2 is the SU(2)
index and $\a$=1,...,4). Note that $\Theta_{\alpha}^{i}$ is a so called
$SU(2)$ Majorana--Weyl spinor (see \cite{kt} for the details).

Worldline superspace is also assumed to be a harmonic (d=1, n=4)
superspace para\-met\-riz\-ed by
\begin{equation}
(\tau,\eta^{i},\bar\eta_{i},u^{\pm}_{i}),
\end{equation}
 and its analytic subspace being parametrized by
\begin{equation}\label{56}
(\tau_{A},\eta^{+},\bar\eta^{+},u^{\pm}_{i})
\end{equation}
\begin{equation}\label{49.}
\tau_{A}=\tau+i\eta^{(i}{\bar{\eta}}^{j)}u^{+}_{i}u^{-}_{j}
\end{equation}
$$\eta^{\pm}=\eta^{i}u^{\pm}_{i},\ \ \ \ \
\bar\eta^{\pm}={\bar{\eta}}^{i}u^{\pm}_{i}.$$

The double analyticity principle claims that the coordinates of
the analytic superspace  \p{53} depend only  on the
analytic superspace coordinates \p{56} of the worldline.

In this respect we note that, as it was demonstrated
in \cite{ds}, the
target space and worldline $SU(2)/U(1)$ harmonic coordinates can be
identified.
This means that the embedding of the worldline harmonic superspace
into the target
harmonic superspace is realized in such a way that the harmonic sectors
$SU(2)/U(1)$ coincide.
$$
X^{m}_{A}=X^{m}_{A}(\tau_{A},\eta^{+},\bar\eta^{+},u^{\pm}_{i})
$$
$$\Theta^{+}_{\alpha}=\Theta^{+}_{\alpha}(\tau_{A},\eta^{+},\bar\eta^{+},
u^{\pm}_{i})$$
\begin{equation}\label{61}
w^{\pm}_{i}=u^{\pm}_{i}.
\end{equation}

Further, the embedding is specified by the geometrodynamical condition
in the following form \cite{ds}
\begin{equation}\label{52}
D^{++}X^{m}_{A}-i\Theta^{+}\gamma^{m}\Theta^{+}=0
\end{equation}
and by the restriction on the superfield
$\Theta^{+}_{\alpha}$
\begin{equation}\label{53.}
D^{++}\Theta^{+}_{\alpha}=0,
\end{equation}
where $D^{++}$ is a  harmonic derivative in the analytic basis
\begin{equation}\label{51.}
D^{++}=u^{+i}{{\partial}\over{\partial{u^{-i}}}}+i\eta^{+}{\bar{\eta}}^{+}
{{\partial}\over{\partial{\tau_{A}}}}.
\end{equation}

To get \p{52} and \p{53.} as consequences of superparticle dynamics the
authors of \cite{ds} proposed
the action
$$S=-i\int\,du\,d{\tau_{A}}\,d\eta^{+}\,d{\bar{\eta}}^{+}[P^{\gamma\delta}
(D^{++}X_{A\gamma\delta}-i\Theta^{+}_{\gamma}\Theta^{+}_{\delta})$$
\begin{equation}\label{54}
+P^{-\alpha}D^{++}\Theta^{+}_{\alpha}]
\end{equation}
which contains two Lagrange multiplier superfields $P_{m}$ and
$P^{-\alpha}$.

A superfield action {\it a la} Ferber--Shirafuji looks as follows
$$S=-i\int\,du\,d{\tau_{A}}\,d\eta^{+}\,d{\bar{\eta}}^{+}[\varepsilon^{\alpha
\beta\gamma\delta}\Lambda^{i}_{\alpha}\Lambda_{\beta{i}}(D^{++}X_{A\gamma
\delta}-i\Theta^{+}_{\gamma}\Theta^{+}_{\delta})$$
\begin{equation}\label{55}
+P^{-\alpha}D^{++}\Theta^{+}_{\alpha}],
\end{equation}
where $\Lambda^{i}_{\alpha}=\Lambda^{i}_{\alpha}(\tau_{A},\eta^{+},
\bar\eta^{+},u^{\pm}_{i})$ are commuting
analytic spinor superfields.

The action \p{55} possesses
N=1 global target space supersymmetry
\begin{equation}
\delta{X^{m}_{A}}=2i\varepsilon^{-}_{\alpha}(\gamma^{m})^{\alpha\beta}\Theta^
{+}_{\beta},\ \ \ \ \ \delta\Theta^{+}_{\alpha}=\varepsilon^{+}_{\alpha}
\end{equation}
$$\delta{P^{-\alpha}}=-2i\varepsilon^{\alpha\beta\gamma\delta}\Lambda^{i}_
{\beta}\Lambda_{i\gamma}\varepsilon^{j}_{\delta}u^{-}_{j}$$
with an odd constant parameter
$\varepsilon^{i}_{\alpha}=u^{+i}\varepsilon^{-}_{\alpha}+u^{-i}\varepsilon^
{+}_{\alpha}$. Eq. \p{55} also has n=4 local worldline supersymmetry
with a local $SU(2)$ automorphism group, bosonic superfield symmetry
\begin{equation}\label{68}
\delta{X_{A\gamma\delta}}={\cal{B}}(\tau_{A},\eta^{+},{\bar{\eta}}^{+},u)
\Lambda^{i}_{\gamma}\Lambda_{\delta{i}},
\end{equation}
and another local $SU(2)$ symmetry which acts on
$\Lambda^{i}_{\alpha}$
\begin{equation}\label{69}
{\Lambda^{'}}^{i}_{\alpha}={U^{i}}_{j}\Lambda^{j}_{\alpha}.
\end{equation}

The equations of motion derived from the action \p{55} have the following
form:
\begin{equation}\label{t}
D^{++}\Theta^{+}_{\alpha}=0
\end{equation}
\begin{equation}\label{g}
\varepsilon^{\alpha\beta\gamma\delta}\Lambda_{\beta{i}}(D^{++}X_{A\gamma
\delta}-i\Theta^{+}_{\gamma}\Theta^{+}_{\delta})=0
\end{equation}
\begin{equation}\label{l}
\varepsilon^{\alpha\beta\gamma\delta}D^{++}(\Lambda^{i}_{\alpha}
\Lambda_{\beta{i}})=0
\end{equation}
\begin{equation}\label{p}
i\varepsilon^{\alpha\beta\gamma\delta}\Lambda^{i}_{\alpha}
\Lambda_{\beta{i}}\Theta^{+}_{\gamma}+D^{++}P^{-\delta}=0
\end{equation}

The solutions to the equations \p{t}, \p{g} and \p{l} are, respectively:
\begin{equation}
\Theta^{+}_{\alpha}=\theta^{i}_{\alpha}u^{+}_{i}+\eta^{+}\tilde{\lambda}
_{\alpha}+
{\bar{\eta}}^{+}{\bar{\tilde{\lambda}}}_{\alpha}-i\eta^{+}{\bar{\eta}}^{+}
{\dot{\theta}}^{i}_{\alpha}u^{-}_{i},
\end{equation}
\begin{equation}
D^{++}X_{A\gamma\delta}-i\Theta^{+}_{\gamma}\Theta^{+}_{\delta}=
\xi(\tau_{A},\eta^{+},{\bar{\eta}}^{+},u)\Lambda^{i}_{\gamma}
\Lambda_{\delta{i}}
\end{equation}
and
\begin{equation}
\varepsilon^{\alpha\beta\gamma\delta}\Lambda^{i}_{\alpha}
\Lambda_{\beta{i}}=
\varepsilon^{\alpha\beta\gamma\delta}\lambda^{i}_{\alpha}
\lambda_{\beta{i}}= const,
\end{equation}
where $\lambda^{i}_{\alpha}=\Lambda^{i}_{\alpha}|_{\eta=0}$.

Using the additional symmetry \p{68} we can choose the gauge $\xi=0$ in
which the geometrodynamical condition \p{g} has the standard form
\begin{equation}
D^{++}X_{A\gamma\delta}-i\Theta^{+}_{\gamma}\Theta^{+}_{\delta}=0.
\end{equation}

To find the solution to eq. \p{p} let us  present $P^{-\alpha}$ in the
component form
\begin{equation}\label{78}
P^{-\alpha}=\Phi^{-\alpha}+\eta^{+}\zeta^{--\alpha}+{\bar{\eta}}^{+}{\bar
{\zeta}}^{--\alpha}+i\eta^{+}{\bar{\eta}}^{+}\psi^{---\alpha}.
\end{equation}
Substituting \p{78} and \p{l} into \p{p} and expending the latter in the
powers of $\eta^{+}$, $\bar\eta^{+}$ we obtain the following component
equations and their solutions:
$$D^{++}\psi^{---\delta}=-
\varepsilon^{\alpha\beta\gamma\delta}\lambda^{i}_{\alpha}
\lambda_{\beta{i}}{\dot{\theta}}^{i}_{\alpha}u^{-}_{i}\Rightarrow$$
\begin{equation}\label{79}
\cases{D^{++}\psi^{---\delta}=0~~~~~\Rightarrow~~~~~ \psi^{---\delta}=0 \cr
\varepsilon^{\alpha\beta\gamma\delta}\lambda^{i}_{\alpha}
\lambda_{\beta{i}}{\dot{\theta}}^{i}_{\alpha}u^{-}_{i}=0 \cr}
\end{equation}

\vspace{1cm}
$$D^{++}\zeta^{--\delta}=-i\varepsilon^{\alpha\beta\gamma\delta}\lambda^{i}
_{\alpha}\lambda_{\beta{i}}{\tilde{\lambda}}_{\gamma}\Rightarrow$$
\begin{equation}\label{80}
\cases{D^{++}\zeta^{--\delta}=0\Rightarrow \zeta^{--\delta}=0 \cr
\varepsilon^{\alpha\beta\gamma\delta}\lambda^{i}_{\alpha}
\lambda_{\beta{i}}{\tilde{\lambda}}_{\gamma}=0 \cr}
\end{equation}

\vspace{1cm}
$$D^{++}\Phi^{-\delta}=-i\varepsilon^{\alpha\beta\gamma\delta}\lambda^{i}
_{\alpha}\lambda_{\beta{i}}\theta^{k}_{\gamma}u^{+}_{k}\Rightarrow$$
\begin{equation}\label{81}
\cases{\Phi^{-\delta}=-i\varepsilon^{\alpha\beta\gamma\delta}\lambda^{i}
_{\alpha}\lambda_{\beta{i}}\theta^{k}_{\gamma}u^{-}_{k}\cr
{\dot{\Phi}}^{-\delta}=0 \cr}
\end{equation}

The relation between the superpartner ${\tilde{\lambda}}_{\alpha}$ of the
Grassmann coordinate $\theta^{i}_{\alpha}$ and $\lambda^{i}_{\alpha}$ is
established with the help of the second equation in \p{80}
\begin{equation}\label{83}
{\tilde{\lambda}}_{\alpha}=a_{i}(\tau_{A})\lambda^{i}_{\alpha}
\qquad
{\bar{\tilde{\lambda}}}_{\alpha}={\bar{a}}_{i}\lambda^{i}_{\alpha}.
\end{equation}

Using the local $SU(2)$ transformations of  worldline n=4
SUSY it is possible to choose the gauge $a_{1}=a$, ${\bar{a}}_{2}=ia$ and
identify ${\tilde{\lambda}}_{\alpha},~{\bar{\tilde{\lambda}}}_{\alpha}$
with the doublet $a\lambda^{i}_{\alpha}$ of the unbroken local $SU(2)$
\p{69}.
Then, as it is easy to see, the geometrodynamical condition \p{g}
reproduces at the component level the solution to the mass shell
condition
\begin{equation}
{\dot{x}}^{m}-2i{\dot{\theta}}^{i}\gamma^{m}\theta_{i}=2a^2\l^{i}
\gamma^{m}\l_{i},
\end{equation}
and upon redefining the fields and eliminating auxiliary variables one
arrives at the conventional formulation of the D=6, N=1 massless
superparticle.

Herein we will not consider a complete supertwistor form of the action
\p{55} since the definition of the supertwistor associated with the
harmonic D=6, N=1 superspace requires additional studies.

\section{Conclusion}

We have considered the formulation of N=1 superparticle mechanics
in D=3,4 and 6,
which is manifestly supersymmetric on the worldline and in the target
superspace. For the construction of the action we have used only
geometrical objects that characterize the properties of embedding
the worldline superspace into the target superspace,
such as target superspace
coordinates of the superparticle and twistor components. The action does
not contain the Lagrange multipliers which may cause the problem of
infinite reducible symmetries, and, in fact, is a worldline superfield
generalization of the supertwistor description of superparticle dynamics
\cite{fer,shir,ced}.

The generalization of the present formulation to the case of N=1, D=10
superparticles and D=3,4,6 and 10 superstrings turned out to be not
straightforward and revealed rather interesting and deep connection of
the twistor--like formulation of super--p--branes with the geometrical
\cite{geo} and the group--manifold \cite{rheo} approach. This problems are
considered elsewhere \cite{bpstv,bsv}.

\bigskip
{\bf Acknowledgements}.
D.S. thanks Paolo Pasti and Mario Tonin for fruitful discussion.
This work was supported  in  part  by  the
International Science Foundation under the grant N RY 9000,
 by the State  Committee  for  Science  and  Technology  of
Ukraine under the Grant N 2/100 and
by the INTAS grants 93--127, 93--493, 93--633.


\newpage
\begin{thebibliography}{99}
\bibitem{stv}
Sorokin D.,  Tkach V. and Volkov D. V. 1989 {\sl Mod. Phys. Lett. A} {\bf 4}
901.
\bibitem{vz}
Volkov D. V.  and  Zheltukhin A. 1988  {\sl JETP Lett.}
{\bf 48} 61; 1989
{\sl Letters in Math. Phys.} {\bf 17} 141;
{\sl Nucl. Phys. B} 1990 {\bf 335}  723.
\bibitem{stvz}
Sorokin D., Tkach V., Volkov D. V. and Zheltukhin A. 1989 {\sl Phys. Lett.
B}
{\bf 216} 302.
\bibitem{sp}
Sorokin D. P.  {\sl Fortshr. Phys.} 1990 {\bf 38} 923.
\bibitem{ds}
Delduc F.  and Sokatchev E. 1992 {\sl Class. Quantum Grav.} {\bf 9} 361.
\bibitem{ps}
Pashnev A.  and  Sorokin D. 1993 {\sl Class. Quantum Grav.}
{\bf 10}  625.
\bibitem{gs92}
Galperin A. and Sokatchev E.  1993 {\sl Phys.
Rev. D} {\bf 46} 714.
\bibitem{hsstr}
Berkovits N. 1990 {\sl Phys.  Lett. B} {\bf 241} 497;\\
Tonin M. 1991 {\sl Phys. Lett. B} {\bf 266} 312; 1992 {\sl Int. J.
Mod. Phys A} {\bf 7} 613;\\
Aoyama S.  Pasti P.  and Tonin M. 1992
{\sl Phys.  Lett. B} {\bf 283} 213;\\
Delduc F.  Galperin A. Howe P.  and Sokatchev E. 1992 {\sl Phys. Rev. D}
{\bf 47} 587.
\bibitem{hsstr2}
 Ivanov E.  and Kapustnikov A.  1991 {\sl Phys. Lett. B} {\bf 267}
175; \\
Delduc F.   Ivanov E. and Sokatchev E. 1992 {\sl Nucl.  Phys. B}
{\bf 384} 334.
\bibitem{chip}
Chikalov V. and Pashnev A. 1993
{\sl Mod.  Phys.  Lett. A} {\bf 8} 285; Preprint ICTP, 1993.
\bibitem{gs93}
 Galperin A. and Sokatchev E. 1993 {\sl Phys. Rev.} {\bf 48} 4810.
\bibitem{tp93}
Pasti P.  and Tonin M. 1994 {\sl Nucl. Phys. B} {\bf 418}
337;\\
Bergshoeff E. and Sezgin E. 1994 {\sl Nucl. Phys. B} {\bf 422} 329.
\bibitem{al}
De Azcarraga J. and Lukierski J. 1982 {\sl Phys. Lett. B} {\bf 113} 170.
\bibitem{sig}
Siegel W. 1983 {\sl Phys. Lett. B} {\bf 128} 397.
\bibitem{gsw}
Green M. Schwarz J. and Witten E. Superstring Theory, CUP, 1987.
\bibitem{spinsup}
 Gates S. J. Jr.  and Nishino H. 1986 {\sl Class. Quantum
Grav.} {\bf 3}  391.\\
 Kowalski-Glikman J. 1986 {\sl Phys.
Lett.} {\bf 180} 358.\\
  Brooks R.  Muhammed  F. and
 Gates S. J. Jr. 1986 {\sl Class.
Quantum Grav.} {\bf 3} 745.\\
Brooks R. 1987 {\sl Phys.  Lett. B} {\bf 186}  313.\\
Kowalski-Glikman J.  van Holten J.  W.  Aoyama S.  and
Lukierski J. {\sl Phys. Lett. B} {\bf  201}  487.\\
Kavalov and  Mkrtchyan R. L.  Spinning superparticles.  Preprint
Yer.PhI 1068(31)-88, Yerevan, 1988 (unpublished).\\
Fisch J. M. L.  1989 {\sl Phys. Lett. B} {\bf 219} 71.
\bibitem{bh}
Bandos I. A.  1990 {\sl Sov.  J.  Nucl.  Phys.} {\bf 51}
 906; 1990 {\sl JETP.  Lett.} {\bf 52} 205.
\bibitem{bzm}
 Bandos I. A.   and Zheltukhin A. A. 1991
 {\sl JETP. Lett.} {\bf 54} 421; {\sl Ibid} 1992 {\bf 55}  81;
 1992 {\sl Phys. Lett. B} {\bf 288} 77;
 1993 {\sl Int. J. Mod. Phys. A} {\bf 8} 1081;
 1992 {\sl Sov. J. Nucl. Phys.} {\bf 56} 198.
 Preprint {\sl IC/92/422} ICTP, Trieste, 1992,
{\sl Sov. J. Elem. Part. Atom. Nucl.} (in press).
 Preprint {\sl DFPD/94/TH/35}, Padova, 1994, {\bf hep-th/9405113}, {\sl
Class. Quantum. Grav.} (in press).
\bibitem{bpstv}
Bandos I. Pasti  Sorokin P.  D. Tonin M. and  Volkov D.
Prerint DFPD 95/TH/02, January 1995, hep-th/9501113.
\bibitem{bsv}
Bandos I. Sorokin  D.  and  Volkov D. Preprint DFPD 95/TH/06, February
1995.
\bibitem{bv}
Batalin I. A. and Vilkovisky G. A. 1983
{\sl Phys. Rev. D} {\bf 28}  2567.
\bibitem{infred}
Bergshoeff E. Kallosh R. and  van
Proeyen A. 1990 {\sl Phys. Lett. B} {\bf 251} 128;\\
 Micovi\'c A.  R\~ocek M.  Siegel W.
et. al. 1990 {\sl Phys. Lett. B} {\bf B235} 106.
\bibitem{fer}
Ferber A. 1977 {\sl Nucl. Phys. B} {\bf 132} 55.
\bibitem{shir}
Shirafuji T. 1983 {\sl Progr. Theor. Phys.} {\bf 70} 18.
\bibitem{ced}
Bengtsson A.K.H. Bengtsson I.  Cederwall M. and  Linden N. 1987 {\sl
Phys. Rev. D} {\bf 36}  1766;\\
 Bengtsson I. and  Cederwall M. 1988 {\sl Nucl. Phys. B} {\bf 30}2 104.
\bibitem{13sp}
Ogievetsky V. and Sokatchev E. 1978 {\sl Phys. Lett. B} {\bf 79} 222.
\bibitem{gikos}
Galperin A. Ivanov E. Kalitzin S. Ogievetsky V. and Sokatchev E. 1984 {\sl
Class. Quantum Grav.} {\bf 1} 469.
\bibitem{15sp}
Galperin A. Ivanov E. Kalitzin S. Ogievetsky V. and Sokatchev E. 1987
{\sl Class. Quantum Grav.} {\bf 4} 1255.
\bibitem{16sp}
Sokatchev E. 1988 {\sl Class. Quantum Grav.} {\bf 5} 1459.
\bibitem{bs}
 Casalbuoni R. 1976 {\sl Nuovo Cimento} {\bf A35} 377;\\
 Pashnev A. and Volkov D. V. 1980 {\sl Teor. Mat. Fiz.} {\bf 44} 310.\\
Brink L. and Schwarz J. 1981 {\sl Phys. Lett.} {\bf B100} 310.
\bibitem{bnsv}
 Bandos I.  Nurmagambetov A.  Sorokin D.  and  Volkov D. 1994
{\sl Sov. JETP Lett.} {\bf 60} 621.
\bibitem{kt}
Howe P. Sierra G. and Townsend P. 1993 {\sl Nucl. Phys. B} {\bf 221}
331;\\
Kugo T. and Townsend P. 1983 {\sl Nucl. Phys. B} {\bf 221} 357.
\bibitem{geo}
Lund F.  and Regge T. 1976 {\sl Phys. Rev. D} {\bf 14} 1524.\\
Omnes R. 1979 {\sl Nucl. Phys. B} {\bf 149} 269.\\
 Barbashov B. M. and Nesterenko V. V.  {\sl Relativistic string model in
hadron physics}, Moscow, Energoatomizdat, 1987 (and refs. therein).\\
Zheltukhin  A. 1981 {\sl Sov. J. Nucl. Phys.} {\bf 33} 1723; 1982 {\sl Theor.
Math. Phys.} {\bf 52} 73;
1982 {\sl Phys. Lett. B} {\bf 116} 147; 1983 {\sl Theor. Math. Phys.}
{\bf 56} 230.
\bibitem{rheo}
 Nieman Y. and  Regge T. 1978 {\sl Phys. Lett. B} {\bf 74} 31;
1978 {\sl Revista del Nuovo Cim.}
{\bf 1} 1;\\
D'Auria R.  Fr\`e P. and T. Regge T. 1980 {\sl Revista del Nuovo Cim.}
{\bf 3} 1;\\
Castellani L.  D'Auria R. Fr\`e P. ``Supergravity and superstrings, a
geometric perspective'', World Scientific, Singapore, 1991 (and
references therein).
\end{thebibliography}
\end{document}